\documentclass[useAMS,usenatbib]{mn2e}
\usepackage{natbib}
\usepackage{aas_macros}
\usepackage{graphicx}
\usepackage{amssymb}
\usepackage{color}

\begin{document}
\date{}

\pagerange{\pageref{firstpage}--\pageref{lastpage}} \pubyear{2002}

\title{Nitrogen depletion in field red giants: mixing during the He flash?}
\author[T. Masseron et al.]{T. Masseron$^{1}$\thanks{E-mail:                 
tpm40@ast.cam.ac.uk},  N. Lagarde$^{2,3}$, A. Miglio$^{3,4}$, Y. Elsworth$^{3,4}$, and G. Gilmore$^{1}$\\
$^{1}$Institute of Astronomy, Madingley Road, Cambridge  CB3 0HA, UK\\
$^{2}$Institut UTINAM, CNRS UMR6213, Universit\'e de Franche-Comt\'e, OSU THETA Franche-Comt\'e-Bourgogne, \\
\hspace{2cm} Observatoire de Besan\c con, BP 1615, 25010 Besan\c con Cedex, France\\
$^{3}$School of Physics and Astronomy, University of Birmingham, Edgbaston, Birmigham, UK\\
$^{4}$Stellar Astrophysics Centre, Department of Physics and Astronomy, Aarhus 
University, Ny Munkegade 120, DK-8000 Aarhus C, Denmark}                                                 
\maketitle

\begin{abstract}
We combine simultaneous constraints on stellar evolutionary status from asteroseismology, and on nitrogen abundances 
derived from large spectroscopic surveys, to follow nitrogen surface abundances all along the evolution of a low-mass
 star, comparing model expectations with data. After testing and calibrating the observed yields from the APOGEE survey,
 we first show that nitrogen surface abundances follow the expected trend after the first dredge-up occurred, i.e.
 that the more massive is the star the more nitrogen is enhanced. Moreover, the behaviour of nitrogen data along the
 evolution confirms the existence of non-canonical extra-mixing on the RGB for all low-mass stars in the field. But
 more surprisingly, the data indicate that nitrogen has been depleted between the RGB tip and the red clump. This may
 suggest that some nitrogen has been burnt near or at the He flash episode. \\
\end{abstract}

\label{firstpage}

\begin{keywords}
Galaxy: abundances -- stars: evolution -- stars: abundances
\end{keywords}

\section{Introduction}
Nucleosynthesis reactions in low-mass giant stars ($\rm\sim 1-3 M_\odot$) modify relatively few surface element abundances. However, among those elements carbon and nitrogen can be quite easily measured in stellar atmospheres via spectroscopic techniques, and thus can be used as stringent constraints for testing our understanding of nucleosynthesis and mixing in evolved stars. 
First dredge-up brings to the stellar surface the nucleosynthetic products of the main sequence burning, while chemical analysis of field stars \citep{Gratton2000} or cluster stars \citep{Lind2009} shows evidence for non-standard mixing along the red giant branch (RGB). Up to now, there has been only one physical mechanism candidate to explain this phenomenon \citep[the thermohaline effect][]{Charbonnel2007}.  More puzzlingly, \citet{Masseron2015} show that extra-mixing has occurred in thin disk stars, but indicate that thick disk stars do not show any evidence for such an extra-mixing process. Although the evolutionary phases of low-mass stars have been well established for decades, the best test for stellar evolution models were bound to observations of a limited number of star clusters \citep[e.g. ][]{Gratton2012} or relatively scattered studies \citep{Tautvaisiene2010}. \\ 
Since the last decade, there has been development of large spectroscopic surveys of Galactic stars such as SDSS\citep{Yanni2009}, Gaia-ESO \citep{Gilmore2012}, APOGEE \citep{Majewski2010} which provide large sample of low-mass giant stars. In parallel, space asteroseismology missions such as CoRoT \citep{Baglin2006} and {\it Kepler} \citep{Gilliland2010} were also launched. The complementary scientific approaches rapidly developed \citep{Pinsonneault2014}. In the end, the large number of stars from large surveys allows us to draw precise trends from homogenously determined chemical abundances, while individual accurate surface gravities improve the calibration.   \\ 
In this work we combine simultaneous input from homogenous N (and C) abundance analyses from APOGEE data \citep[DR12][]{Holtzman2015}, accurate abundances from \citet{Hawkins2016}, and accurate masses from {\it Kepler} light curves \citep{Pinsonneault2014} for a large sample of stars. This leads us to track the evolution of N abundance evolution for low-mass stars in the field.

\section{Data sample and models}\label{sec:data}
\subsection{The sample}
\begin{figure}
\includegraphics[width=6cm,angle=-90]{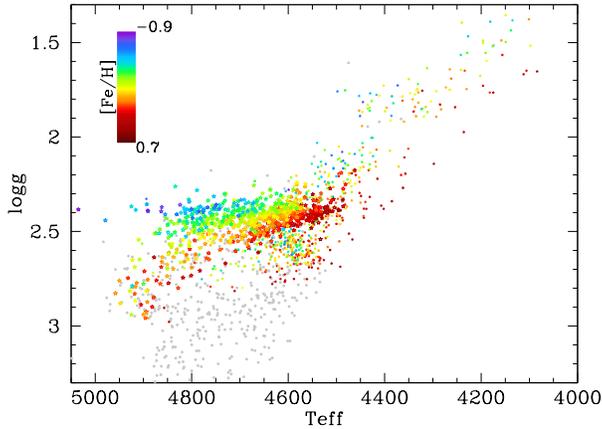}
\caption{ HR diagram of the APOKASC sample color-coded in metallicity. The open symbols are for red clump stars and closed symbols are for RGB stars, while the grey points stand for unclassified stars.}\label{fig:HR_APOKASC}
\end{figure}
\begin{figure}
\includegraphics[width=6cm,angle=-90]{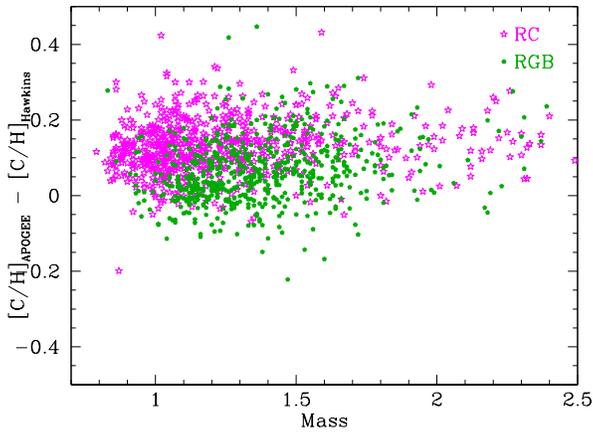}
\includegraphics[width=6cm,angle=-90]{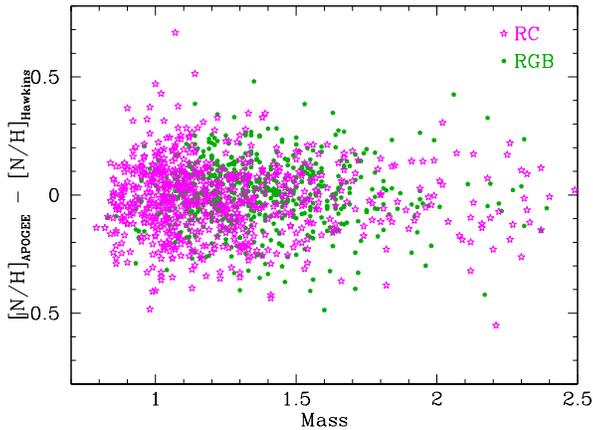}
  \caption{Comparison of C and N abundances in the APOKASC sample stars between the APOGEE DR12 \citep{Holtzman2015} and \citet{Hawkins2016}.}\label{fig:CNspectrum}
\end{figure}
\begin{figure}
\includegraphics[width=6cm,angle=-90]{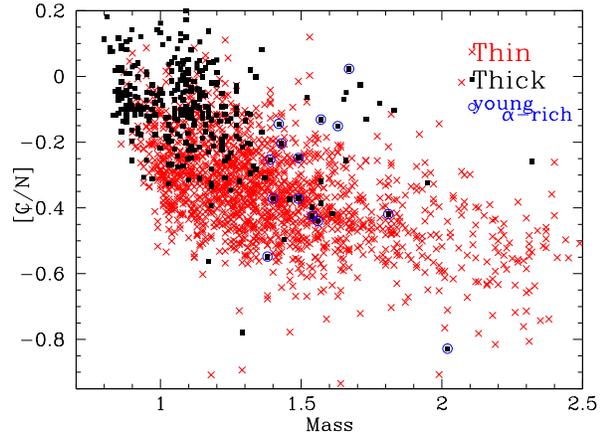}
\includegraphics[width=6cm,angle=-90]{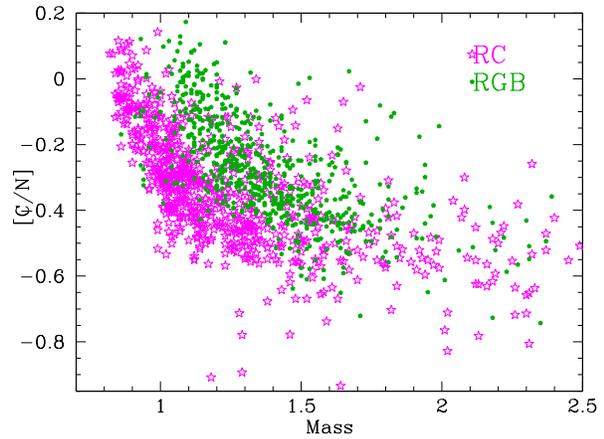}
\includegraphics[width=6cm,angle=-90]{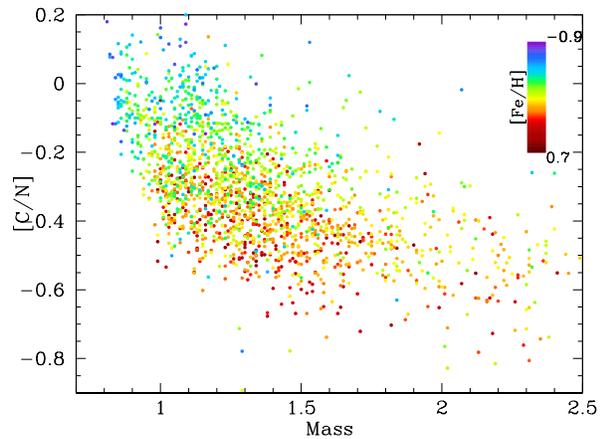}
 \caption{ C/N ratios as a function of mass from the whole APOKASC sample. Upper panel: red crosses are thin disk stars while black squares are thick disk stars. Blue symbols highlight the young $\alpha$-rich stars population as identified by \citet{Martig2015}. Middle panel: magenta stars are red clump stars and green pentagons are RGB stars. Bottom panel: the same stars but color-coded in metallicity. }\label{fig:CNvsMass_APOKASC}
\end{figure}
To build up the sample, we cross match the APOGEE abundance data from \citet{Holtzman2015} with the asteroseismic masses and stellar parameters ($T_{eff}, \log g$)  from \citet{Pinsonneault2014}, as well as the evolutionary status \citep{Elsworth2016}. 
We also took the data for stars in the open cluster NGC6819 from  \citet{Pinsonneault2014}. Fig.~\ref{fig:HR_APOKASC} shows the whole APOKASC sample as used in this work. We divide the sample into three subcategories: ``low RGB'' (RGB stars with $\log g > 2.1$), (``upper RGB'' (RGB stars with $\log g < 2.1$) and ``clump'' stars (as defined by asteroseismology). The separation criteria in $\log g$ has been chosen to approximately match the RGB bump (and thus the beginning of the occurrence of the extra-mixing). First of all, it is noticeable that the sample is mostly comprised of low-RGB and clump stars. Concerning the upper RGB stars, the data are relatively scarce, especially when considering only restricted regimes in metallicity. This observation led us to attempt to extend the sample in order to explore this part of the evolution by some empirical relation (Sec.~\ref{sec:extending}). It is also important to mention that all the stars in this sample are evolved enough to have all undergone their first dredge-up.\\
To distinguish the different Galactic populations, a selection based on the $\alpha$ abundances has been applied such that thick and thin disk stars are empirically distinguished by their $[\alpha/Fe]$ content, such that \[ [\alpha/Fe] \le 0.06 \times [Fe/H] + 0.1 \]

Biases on masses are estimated to be at 10\% or lower level \citep[e.g., see][]{Miglio2012,Brogaard2016,Davies2016,Miglio2016}, and this has virtually no impact on the trends we see in this paper.\\

Because we use C and N abundances from APOGEE DR12 and compare to the stellar evolution predictions, it is important to check their respective accuracy. In particular, based on the observations of the subgiants, \citet{Masseron2015} suspected a systematic offset in N in the APOGEE DR12 data. Indeed, it is crucial to evaluate more precisely the impact of the inconsistency between the stellar parameters of \citet{Pinsonneault2014} and those used by APOGEE DR12 to derive the C and N abundances. Fig.~\ref{fig:CNspectrum} compares the DR12 abundances to the abundances obtained by \citet{Hawkins2016} using \citet{Pinsonneault2014} stellar parameters. This figure shows that there is no significant offset in N between the two independent studies. This suggests that the N offset as observed by \citet{Masseron2015} is dependent on temperature and affects more significantly the subgiant spectroscopic regime than the RGB/clump regime as we study here. Therefore, we do not apply any correction to the N abundances. \\
It is also interesting to note that, while there is no discrepancy in nitrogen, there is a moderate one in carbon.  Indeed, as shown by \citet{Masseron2015}, C measurement based on the CO molecule is more sensitive to stellar parameters, particularly to $\log g$, in contrast to the CN molecule. Consequently, given that \citet{Holtzman2015} highlight a discrepancy on $\log g$ between RGB and clump stars in DR12 data, we expect a discrepancy in C between clump stars and RGB. Therefore, we applied a -0.06 offset to the C abundances for the RGB stars and a -0.12 offset for the clump stars for the whole APOGEE sample, while we have not applied any correction to N. \\

Fig.\ref{fig:CNvsMass_APOKASC} shows the corrected C/N ratios as a function of mass for the whole sample, but color-coded by: a) stellar population origin, b) stellar evolutionary status, and c) metallicity. This figure shows primarily that C/N is anti-correlated with stellar mass, mostly due to the fact that the more massive is the star the more CN-processing occurred. One may also notice in the middle panel that the C/N trend follows two parallel sequences between the RGB and the clump stars. This is due to the mass lost during the RGB phase similarly to what is already observed in clusters \citep{Miglio2012}. We can also directly verify in the top panel that thick disk stars have on average a lower mass than thin disk stars as predicted by \citet{Masseron2015}. However, we also observe in this panel that some stars do not follow the expected pattern. According to \citet{Jofre2016}, those stars which correspond to the "young $\alpha$-rich" population \citep{Martig2015} but are likely to be the result of binary interactions, thus affecting their C/N ratios.
  
Thanks to the high data quality and the large parameter extent of the sample, we can now explore and test those effects by isolating the relevant parameters and comparing them to model expectations. We assume in this paper that the N abundances reflect only internal processes (in other words that the initial composition is negligible compared to the internal production) and thus is consistent with the models (which assume [N/Fe]=0). In contrast, C is enhanced at low metallicity \citep{Masseron2015,Nissen2014}, and thus cannot be consistently compared to the models which generally assume [C/Fe] = 0. This is why we primarily focus here on N abundance for discussions on stellar evolution.

\subsection{Stellar evolution models}
In order to illustrate the standard expectations of the C and N surface abundances, we display along with the data points different stellar evolution models: BaSTI \citep{Pietrinferni2004,Pietrinferni2006}, PARSEC \citep{Bressan2012} and STAREVOL \citep{Lagarde2012}. 

Those theoretical models adopt different prescriptions (e.g. opacities, equation of state, nuclear reaction rates) which provide an indication of the overall uncertainty of the theoretical physical quantities. The variation in the physical inputs notably include the initial composition, for which BaSTI models employ the \citet{Grevesse1993} solar abundances, PARSEC models use the \citet{Caffau2011} values and STAREVOL the \citet{Asplund2005} values except for Ne for which they adopt the value derived from \citet{Cunha2006}. The mixing length parameters are different as well (respectively 2.01, 1.74, and 1.6).\\
All the models include convective core overshooting during the main sequence, but with various efficiencies. In addition, PARSEC models account for overshooting from the convective envelope and atomic diffusion, partially inhibited from the surface convective layers. Concerning rotation, only STAREVOL models have made calculations with and without rotation, using the formalism developed by \citet{Zahn92}, and \citet{MaZa98} \citep[see more details in ][]{Lagarde2012}. They assume a rotational velocity at the zero-age-main-sequence equal to 30\% of the critical velocity for each mass. Moreover, \citet{Lagarde2012}'s models provide a self-consistent prescription of the extra-mixing along the red giant branch up to the early-AGB phase. Indeed, they consider that thermohaline instability develops as long thin fingers with an aspect ratio ($\alpha_{th}$=5) consistent with prescriptions by \citet{Ulrich72} and laboratory experiments \citep{Krish03}. This instability reproduces very well RGB abundance patterns at all metallicities \citep{ChaLag10}. A more detailed and quantitative discussion of those prescriptions on C and N surface abundances can be found in \citet{Salaris2015}. 

\section{Discussion}
\subsection{Constraints on stellar evolution}\label{sec:evol}
\subsubsection{Solar metallicity}
\begin{figure}
\includegraphics[width=6cm,angle=-90]{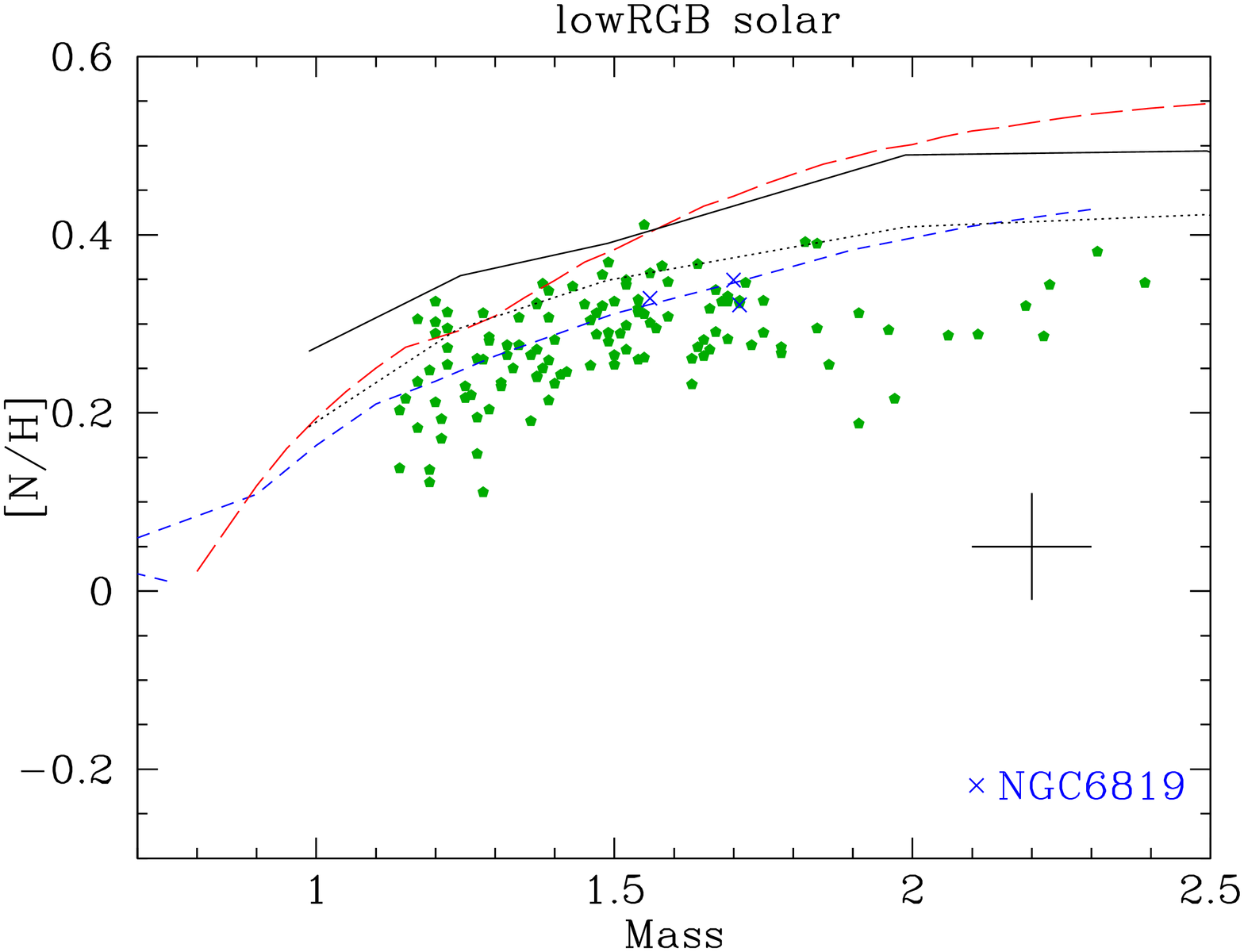}
\includegraphics[width=6cm,angle=-90]{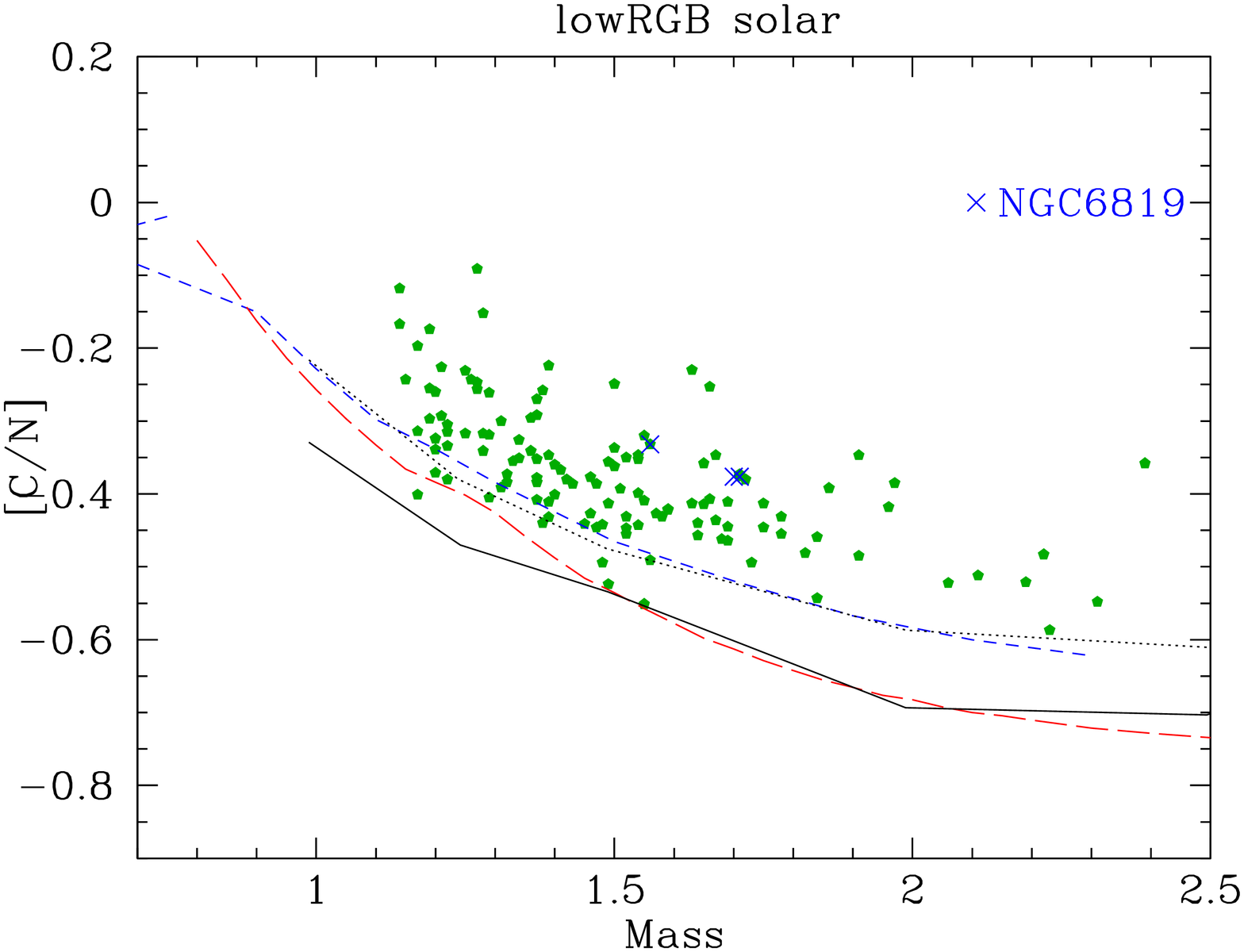}
\caption{[N/H] and [C/N] as a function of mass for the solar metallicity low RGB/post 1$\rm^{st}$ dredge-up stars. Blue crosses also indicate stars belonging to the near-solar metallicity open cluster NGC6819. The models show post first dredge-up values from STAREVOL, with rotation and thermohaline mixing (solid line) and without (dotted line), PARSEC (red long dashed) and BaSTI (blue short dashed).}\label{fig:CNvsMass_lowRGB}
\end{figure}
The first test we made consists in the simplest case to be compared to stellar evolution theory: solar metallicity (by solar metallicity we consider in fact stars for which metallicities correspond to $[Fe/H]=0.0\pm 0.1$ and belong to the thin disk).  In addition, we also consider stars in the open cluster NGC6819 because its metallicity is also about solar and thus are expected to follow the same trends as for field stars. Furthermore, because we select stars with solar metallicity, we assume that their initial composition is also solar (i.e. [C/H]=0 and [N/H]=0). \\ 
\\
 
{\it Low-RGB stars and the first dredge-up phase}\label{sec:rotation}\\
Fig.~\ref{fig:CNvsMass_lowRGB} shows the evolution of the [C/N] ratio and [N/H] for this subsample compared to model expectations. All the models reproduce the same trend for both elements: C is increasingly depleted with increasing mass to the benefit of N which shows a global increasing trend against mass. This leads to the first dredge-up, where H-processed material is diluted in the convective envelope.
 The depth of the dredge-up is expected to be correlated with the mass of the star, hence the observation of higher N and lower C abundances for the higher masses. However, there are various offsets between the models that can be attributed to their various prescriptions. But it is beyond the scope of this paper to compare the models and infer some of their prescriptions. Nevertheless, from the comparison between the data points and the models, we can obtain a relative estimate of the confidence level for our discussion. We can observe from Fig.~\ref{fig:CNvsMass_lowRGB} that all the models agree within $\approx$0.1 dex with the average data points. Therefore, we will assume that the data follow the expected trend as long as the difference between the models and the average data remains within such value.

{\it Clump stars and the core He burning phase}\\
\begin{figure}
\includegraphics[width=6cm,angle=-90]{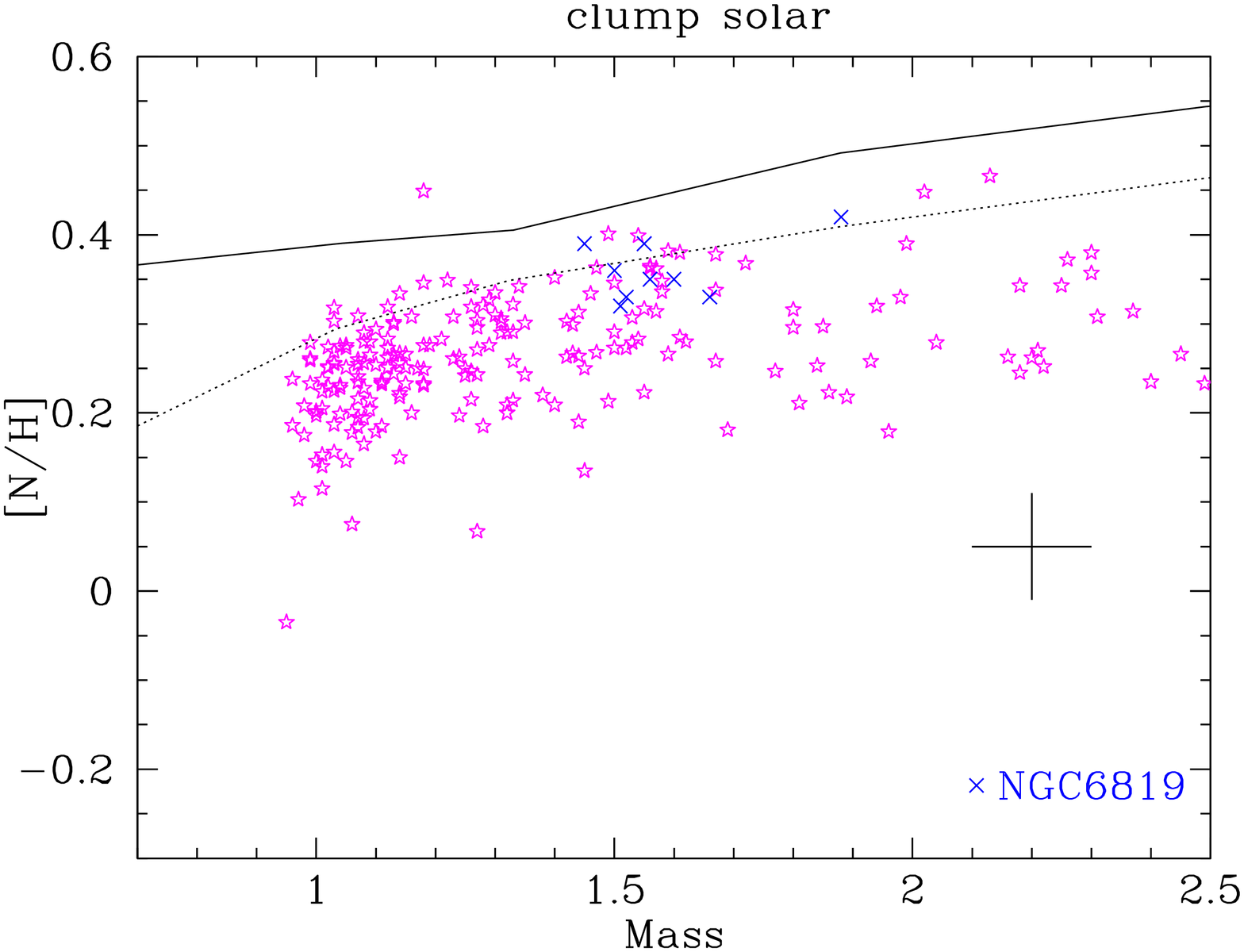}
\includegraphics[width=6cm,angle=-90]{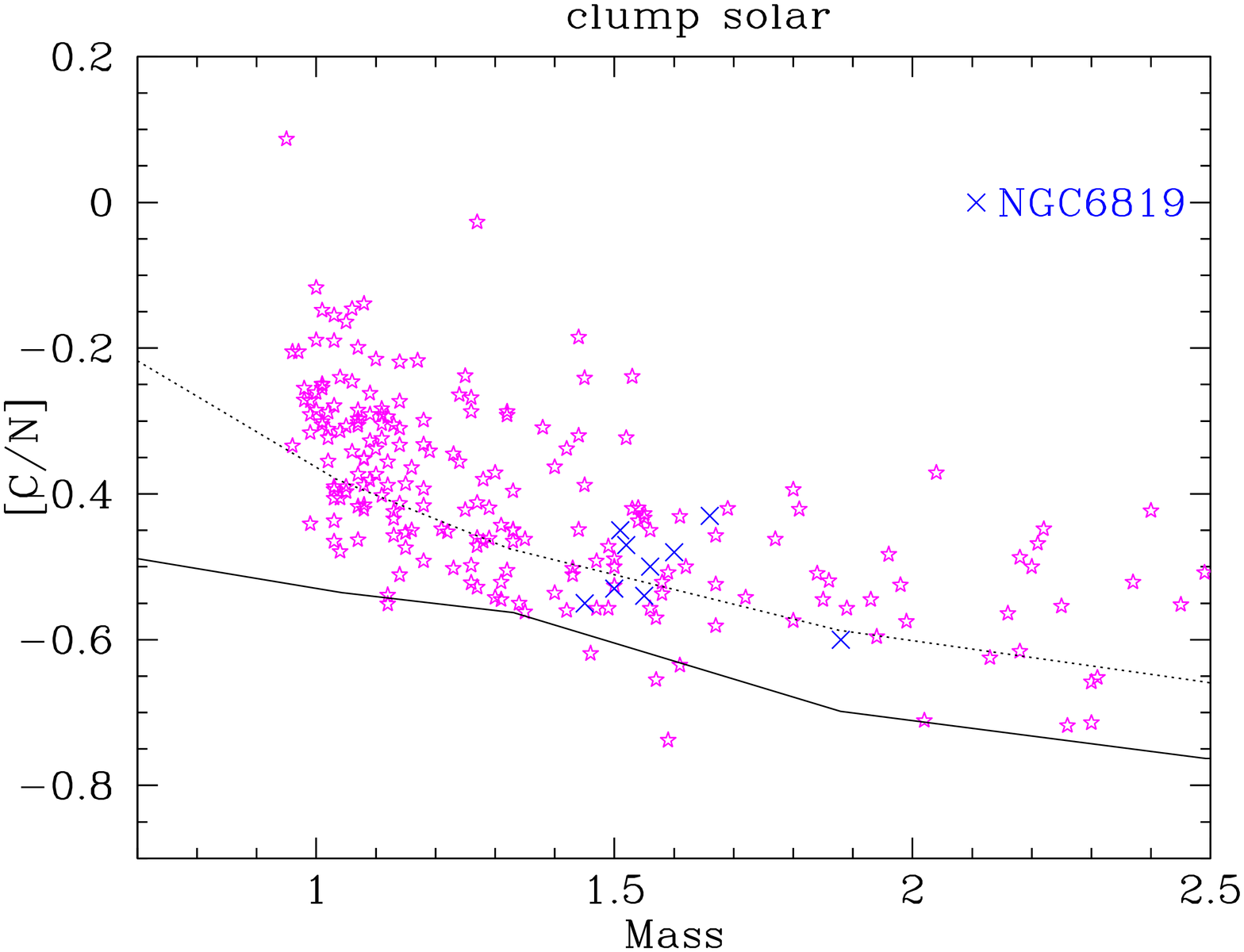}
  \caption{[N/H] and [C/N] as a function of mass for solar metallicity clump stars. Blue crosses also indicate stars belonging to the near-solar metallicity open cluster NGC6819. The models are from \citet{Lagarde2012}, with rotation and thermohaline mixing (solid line) and without (dotted line). }\label{fig:NFevsMass_clump}
\end{figure}
Fig.~\ref{fig:NFevsMass_clump} shows [N/H] and [C/N] as a function of mass for solar metallicity stars while in their core He burning phase. It is expected that non-canonical extra-mixing occurred between the bump and the core He burning phase, further enhancing N (and depleting C). This effect is also expected to be more effective for the lowest mass stars as the model with thermohaline mixing illustrates in Fig.~\ref{fig:NFevsMass_clump}. But the data do not seem to confirm that N has been enhanced at the clump and, more generally almost no change in abundances appear since the first dredge-up values. However, extra-mixing along the RGB has been mostly highlighted in low-metallicity stars. This is what we will verify with our sample in the following section.

\subsubsection{Low metallicity}\label{sec:metallicity}
\begin{figure}
\includegraphics[width=6cm,angle=-90]{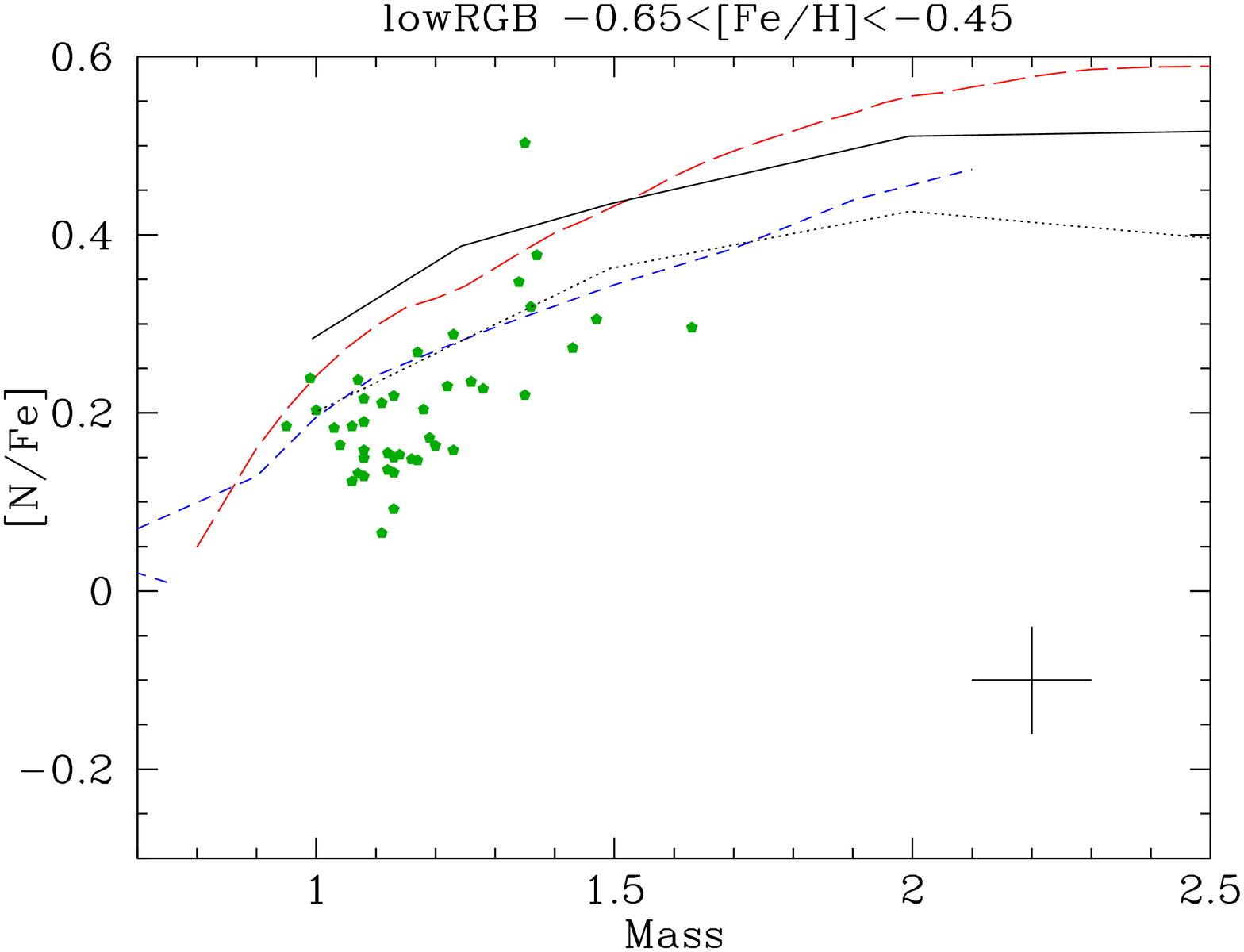}
\includegraphics[width=6cm,angle=-90]{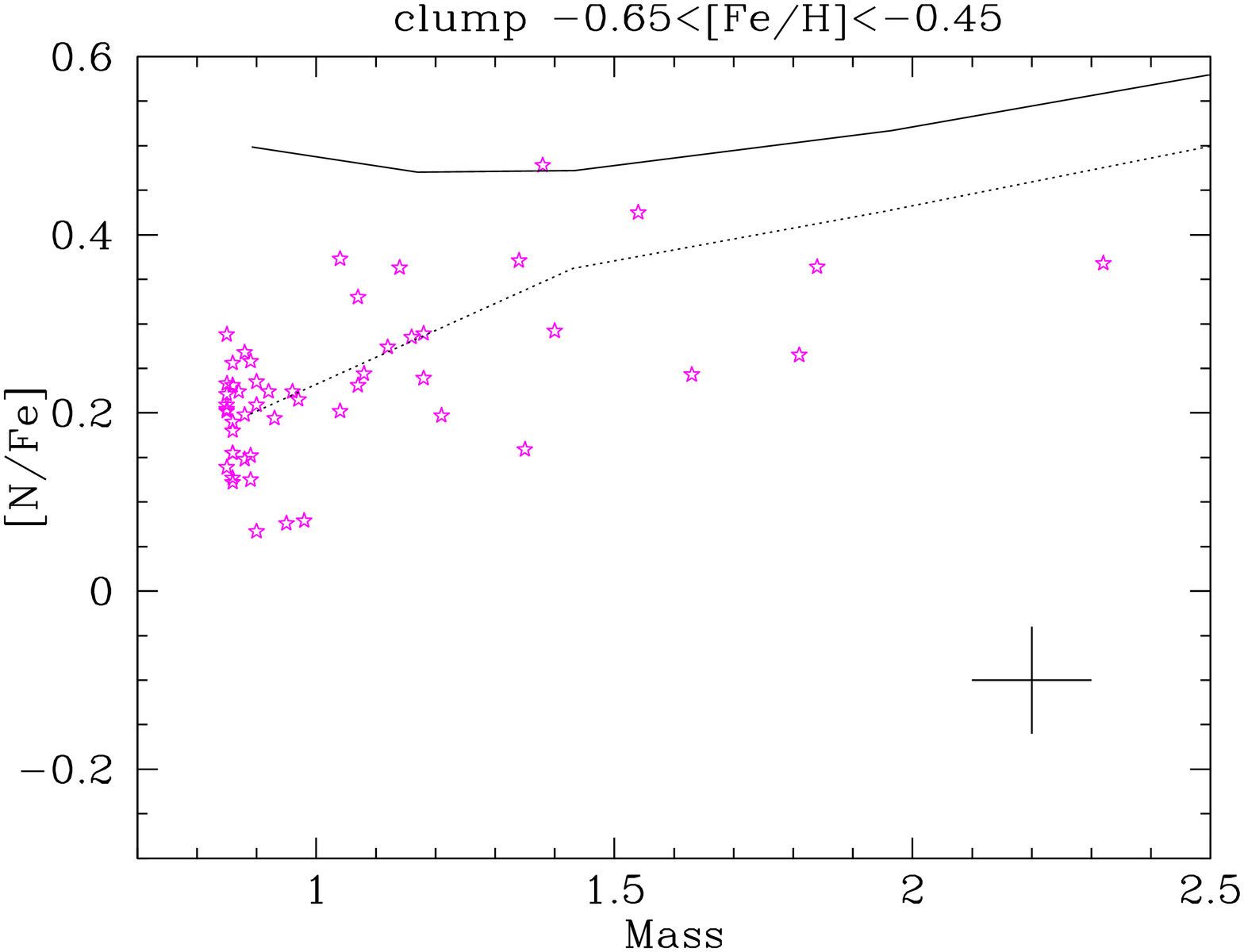}
\caption{[N/Fe] abundance as a function of mass for [Fe/H]$\sim -0.5$ low RGB stars (upper panel) and  [Fe/H]$\sim -0.5$ clump stars (bottom panel). The models are the post first dredge-up values of the Z=0.004 models from STAREVOL \citep{Lagarde2012}, with rotation and thermohaline mixing (solid line) and without (dotted line), PARSEC \citep[red long dashed]{Bressan2012} and BaSTI \citep[blue short dashed]{Salaris2015}.}\label{fig:NFevsMass_Fepoor}
\end{figure}

In this section, we select only low-metallicity stars in the sample such that $\rm [Fe/H]=-0.55\pm0.1$ and that belong to the thin disk population. This metallicity has been chosen to conveniently match the available Z=0.004 models. Fig.~\ref{fig:NFevsMass_Fepoor} shows the related results against the models.   \\

1$\rm^{st}$ dredge-up predictions of the low-metallicity model reproduce relatively satisfactorily N abundance in low-metallicity low-RGB stars (top panel of Fig.~\ref{fig:NFevsMass_Fepoor}), although an offset similar to the solar metallicity case remains related to the model prescriptions. Between the top panel and bottom panel of Fig.~\ref{fig:NFevsMass_Fepoor} non-canonical extra-mixing is expected to have occurred, enhancing N at the surface. This effect is more pronounced as the star is less massive and more metal-poor. This is indeed supported by some very metal-poor stars or globular cluster star observations \citep{Gratton2000,Lind2009}. But no surface abundance change is expected at the He burning phase and all stars should show large N abundance. This appears in contradiction with the observations presented in Fig.~\ref{fig:NFevsMass_Fepoor}. \\
 There are three hypotheses to explain this discrepancy: i) the N data for low-metallicity clump stars are wrong, ii) the extra-mixing did not occur in field stars, or iii) N has been depleted between the end of the H shell burning phase and the He core burning phase. Regarding the first hypothesis, we argue that stellar parameters between RGB and clump stars are too similar to account for a systematic parameter dependant error in the determined abundance (see Fig.~\ref{fig:HR_APOKASC}). We also illustrate in Fig.~\ref{fig:CNspectrum} the absence of any systematic effects due to errors in known stellar parameters. Therefore, we assume that the discrepancy is not related to measurement errors, but is real.    
To discriminate between the two other hypotheses, we need to explore what happened along the upper part of the RGB. Unfortunately, as illustrated in Fig.~\ref{fig:HR_APOKASC} and \ref{fig:CNvsMass_APOKASC}, the APOKASC sample contains too few upper RGB stars to be discriminating. Therefore, in Sec. \ref{sec:extending}, we extend the APOKASC sample with more APOGEE data all along the RGB. \\

\subsection{Extending the sample}\label{sec:extending}
\begin{figure}
\includegraphics[width=6cm,angle=-90]{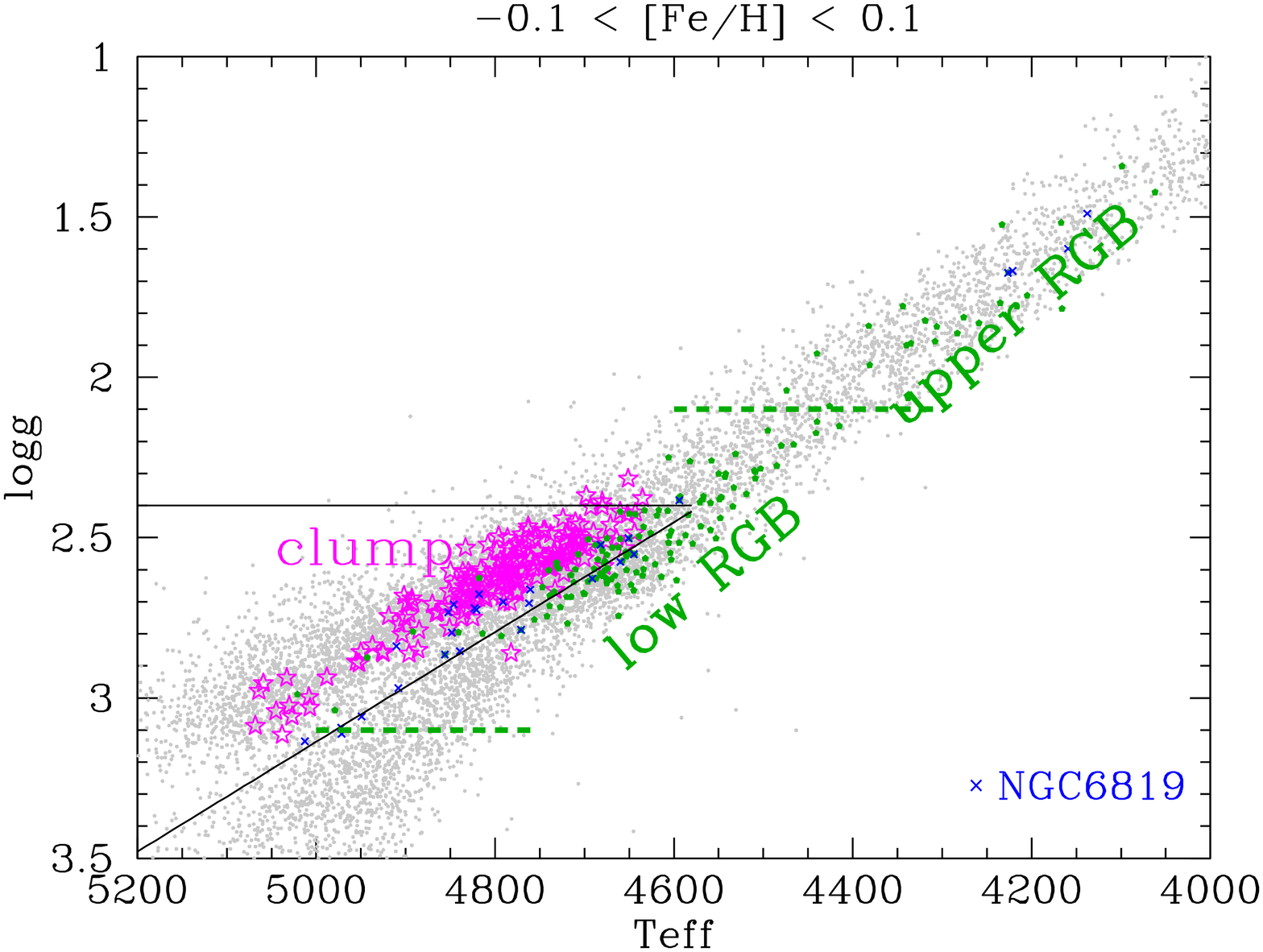}
\includegraphics[width=6cm,angle=-90]{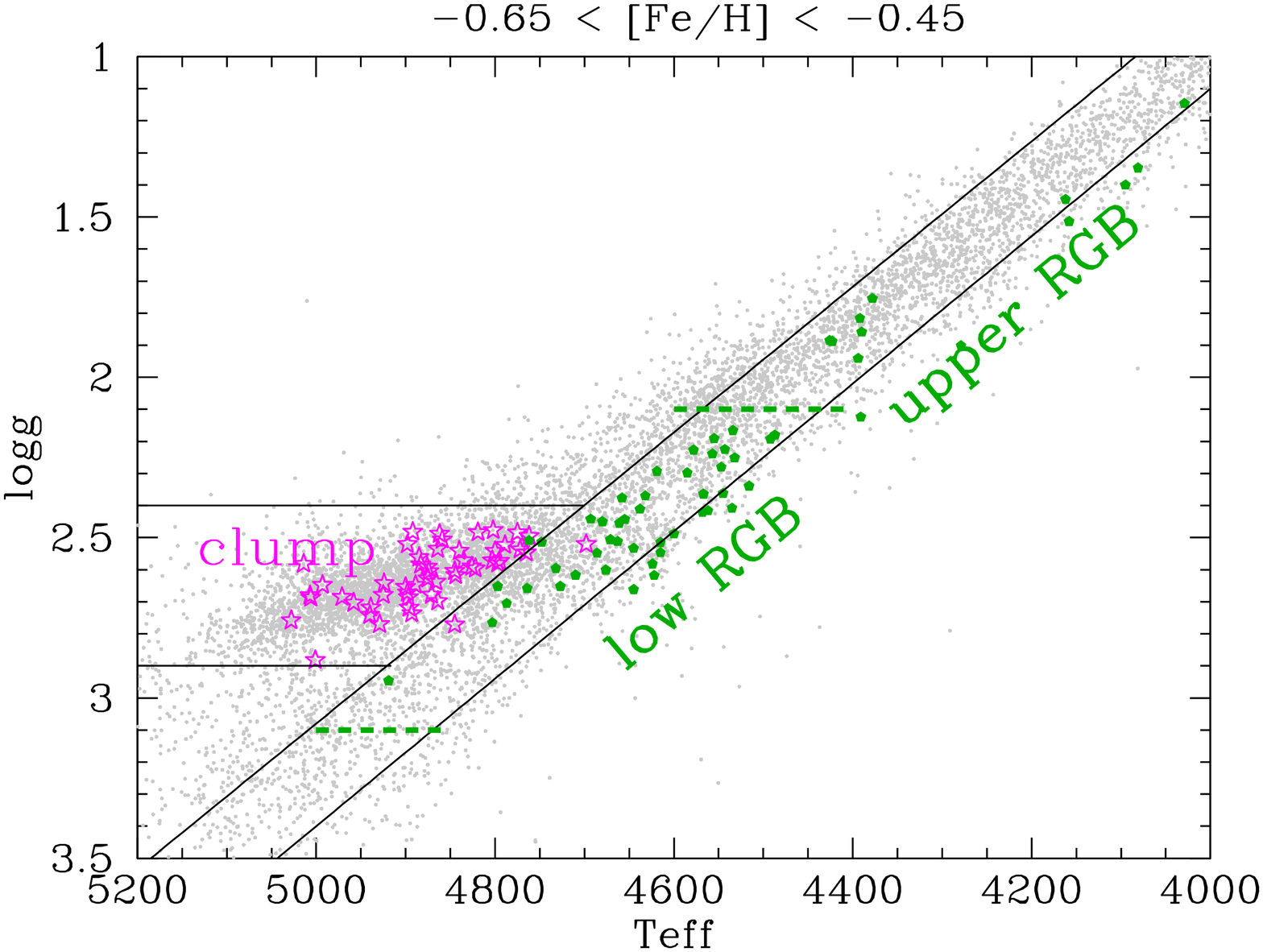}
\caption{HR diagram for two subsets of the APOGEE sample: solar metallicity stars ($-0.1 < [Fe/H] < -0.1$ and stars of the open cluster NGC6819) and low metallicity stars ($-0.65 < [Fe/H] < -0.55$). The coloured points indicate stars which are also part of the APOKASC catalogue, thus with asteroseismic classification RC (magenta stars) and RGB (green pentagons). From that empirical comparison, regions are defined in order to assign an evolutionary status (low RGB, upper RGB and clump) to APOGEE stars. Note that for consistency, we exclusively use the spectroscopic temperature, surface gravity, and metallicity as determined from APOGEE DR12 (even for the stars with asteroseismic targets).}\label{fig:HRfull}
\end{figure}
\begin{figure}
\includegraphics[width=6cm,angle=-90]{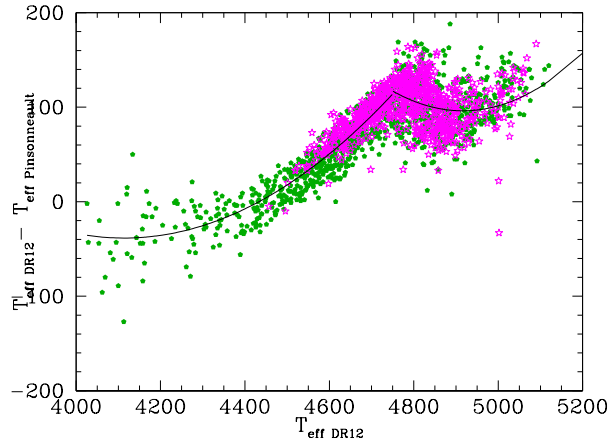}
\caption{Difference in effective temperature between the adopted APOGEE DR12 data and the temperature recommended by \citet{Pinsonneault2014}. The coloured points indicate stars with asteroseismic classification RC (magenta stars) and non-RC (green pentagons). The line shows the empirical temperature correction applied to the APOGEE DR12 data.}\label{fig:Teffcorr}
\end{figure}

We want here to further understand what happened to the N abundances along the RGB sequence, up to the He-core burning phase. Because of the complex interplay between mass, metallicity, and chemical evolution we also disentangle this problem by splitting the sample into specific metallicity bins, solar ([Fe/H]$\sim$ 0.0, i.e. Z=0.014) and low metallicity ([Fe/H]$\sim$-0.55, i.e. Z=0.004). However, as illustrated in Fig.~\ref{fig:HR_APOKASC}, there are too few upper RGB in the APOKASC sample to allow such splitting. Therefore, we decided to extend the APOKASC sample to the whole APOGEE sample. In order to classify the star, we use the information given by the APOKASC sample to derive an empirical selection criteria to separate RGB and clump stars (Fig.~\ref{fig:HRfull}). Naturally, we expect some misclassification and thus some contamination of RGB stars in the clump sample and vice versa. Nevertheless, in the forthcoming discussions, we expect it to be only for stars around 4600K and we will mostly focus on the bulk of stars and their general trend. Additionally, we apply an empirical correction to the effective temperature scale of the APOGEE DR12 sample so that they match the temperatures of \citet{Pinsonneault2014} (Fig.~\ref{fig:Teffcorr}). \\

\subsubsection{Extra-mixing along the RGB and mixing during He flash}\label{sec:Heflash}
\begin{figure}
\includegraphics[width=6cm,angle=-90]{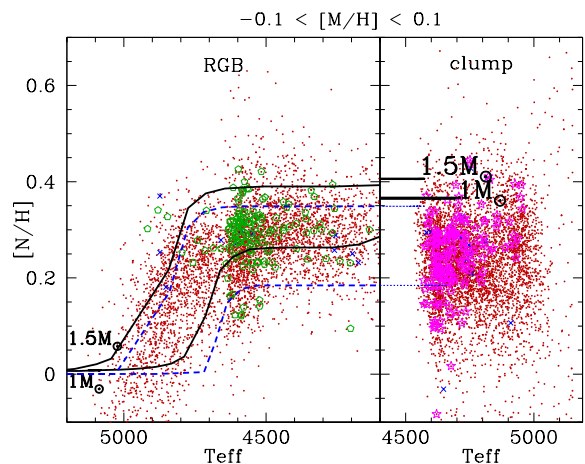}
 \caption{[N/Fe] as a function of effective temperature for the solar metallicity stars of the APOGEE sample. Green pentagons and magenta starred symbols indicate stars present in the APOKASC catalogue. Blue crosses stand for stars belonging to the open cluster NGC6819. Lines represent models with Z=0.014, and 1M$_\odot$ and 1.5M$_\odot$ from \citet{Lagarde2012} with rotation and thermohaline mixing (continuous black) and without (blue dashed line). }\label{fig:NFevsFe_APOGEE_solar}
\end{figure}

\begin{figure}
\includegraphics[width=6cm,angle=-90]{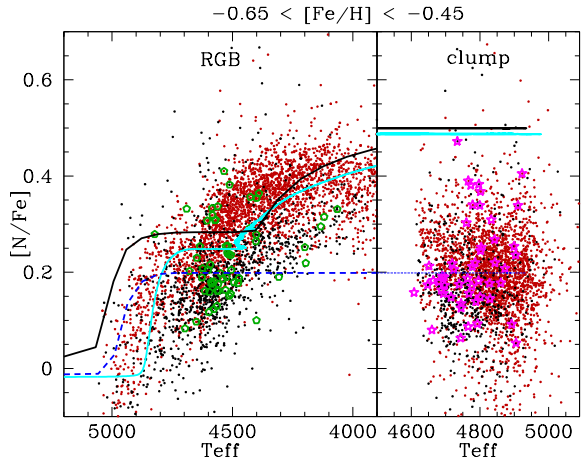}
 \caption{[N/Fe] as a function of effective temperature for low-metallicity stars of the APOGEE sample. Black points are thick disk stars and red points are thin disk stars. Green pentagons and magenta starred symbols indicate stars present in the APOKASC catalogue. Lines represent models of $\rm1M_\odot$, Z=0.004, with enhanced (see text for details) initial C and O and thermohaline mixing only (cyan continuous), with solar scaled CNO abundances and thermohaline + rotation (black continuous), and with solar scaled CNO abundances with neither rotation nor thermohaline mixing (blue dashed). }\label{fig:NFevsFe_APOGEE_Fepoor}
\end{figure}

Fig.~\ref{fig:NFevsFe_APOGEE_solar} shows the N abundance evolution against effective temperature along the RGB and up to the clump for the ``standard solar case'' (i.e. solar metallicity and thin disk composition) with a comparison to model expectations. Although we do not have mass information for that sample, we expect the stars of this subsample mostly to have a main sequence mass of 1 to 1.5M$_\odot$ by reasonably  assuming  a standard IMF distribution and thin disk age.  
We can first observe in this figure that the N abundances increase after the first dredge-up which occurs at $\rm T_{eff}\approx$5000-4800K and then flatten off, consistently with the models. The expected N abundances at the clump stage should be $\approx$0.4 dex above solar, almost independently of the initial mass. We confirm here that red clump stars do not support model expectations at solar metallicity. This increase of N surface abundance at the clump is expected because extra-mixing has occurred over the upper RGB phase since the RGB bump. However, we cannot still confirm the occurrence of the extra-mixing on the upper RGB, because this occurs for the lowest mass stars at $\rm T_{eff} \approx 4000K$ at solar metallicity, beyond the range of observations.\\
At low metallicity, non-canonical extra-mixing occurs at higher $\rm T_{eff}$ and thus can be studied in the APOGEE sample (Fig.~\ref{fig:NFevsFe_APOGEE_Fepoor}).  Indeed, this figure shows that all the stars, whether they belong to the thin or the thick disk, show an increasing N abundance along the RGB. Therefore, this proves that extra-mixing occurs as well in all disk stars whatever is their metallicity, ruling out a suggestion of \citet{Masseron2015}. It is worth recalling here that, in contrast to solar metallicity, it is expected that in the low metallicity regime the stellar population is dominated by $\sim$ 1M$_\odot$ stars (see also Fig.~\ref{fig:CNvsMass_APOKASC}).\\

We also address here the question of the impact of a change in its initial composition on the evolution of the star. In particular thick disk stars are known to be enhanced in O \citep[among $\alpha$-elements][]{Fuhrmann1998} and in C \citep{Nissen2014,Masseron2015}. Hence, we computed a 1M$_\odot$ stellar model consistently with the models of \citet{Lagarde2012}, such that Z=0.004 and [C/Fe]=+0.2 and [O/Fe]=+0.2. While Fig.~\ref{fig:NFevsFe_APOGEE_Fepoor} shows that the net abundances of N have not been significantly affected, the time scale and the occurrence of the first dredge-up and extra-mixing have been boosted up.  In particular, both models predict high N surface abundance when reaching the He core burning phase. Therefore, the change in initial composition does not seem to explain the apparent low N abundances in the clump stars. \\

\begin{figure}
\includegraphics[width=6cm,angle=-90]{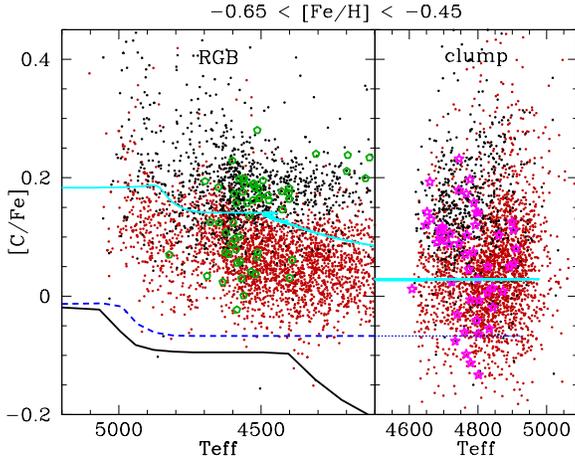}
 \caption{[C/Fe] as a function of effective temperature for low-metallicity stars of the APOGEE sample. Symbols and lines are the same as in Fig.~\ref{fig:NFevsFe_APOGEE_Fepoor}. }\label{fig:CFevsFe_APOGEE_Fepoor}
\end{figure}

Could the low abundance of N at the clump be due to a measurement bias? We have already shown in Sec.~\ref{sec:metallicity} that red clump stars have very similar stellar parameters than RGB stars at $\sim$4600K. Hence, given that similar stellar parameters necessarily provide similar abundances for a given line strength, it is difficult to claim any systematic offset in abundance between those two groups. We further argue here that, if there is still a bias in the N data, this bias must then be continuously increasing towards cooler temperatures. But the only way to reconcile the N abundance at the end of the RGB sequence ($\sim$3900K) with that observed in clump stars would be to have N constant all along the temperature sequence. This would imply that non-canonical extra-mixing on the RGB does not exist in field stars. Knowing that Fig.~\ref{fig:NFevsFe_APOGEE_Fepoor} shows a clear signature of extra-mixing  along the RGB consistently with several literature works \citep[e.g.][]{Gilroy1989,Gratton2000}, we then assume that the N depletion between upper RGB and clump stars is real. \\
Another scenario would consist in invoking the fact that the clump and RGB stars belong to two distinct generations of stars, as it is observed in some globular clusters \citep{Gratton2012}. However, this scenario is very unlikely because, in contrast to clusters, both thin and thick disks have a known extended star formation history.\\
Another possibility consists in considering that N has been destroyed during or after the flash. One suggestion from \citet{Eggleton1968} predicted that a N flash could occur, through the $\rm ^{14}N(\alpha,\gamma)^{18}O$ reaction just before the He flash, but only in low-mass stars when a large enough amount of CNO is present. However, with some updated reaction rates \citet{Couch1972} argued that this was very unlikely. Current stellar evolution models \citep[e.g.][]{Bildsten2012} confirm the findings of \citet{Couch1972}. Furthermore, after a decade of debate, the most recent dynamical study of \citet{Deupree1996} concludes that there is no mixing between the convective envelope and the He burning region. More recent state of the art full hydro simulations by \citet{Mocak2011} go the same direction.\\ 
Another theoretical difficulty for mixing to occur would be to mix material through the H burning shell despite the entropy jump. Nevertheless, such a possibility is expected to happen \citep{Hollowell1990,Schlattl2001,Campbell2008}, although at much lower metallicity where the entropy jump is lower ($\rm [Fe/H]\lesssim -3.0)$. Indeed, such a mixing implies that N meets He at high enough temperature to be detroyed by the $\rm ^{14}N(\alpha,\gamma)^{18}O(\alpha,\gamma)^{22}Ne$ reaction, but also creates lots of $\rm ^{12}C$ from triggering the triple-$\alpha$ reaction, as well as enough free neutrons via proton capture on $\rm ^{12}C$, hence potentially leading to s-process nucleosynthesis \citep{Cruz2013}. However, $\rm ^{12}C$ does not seem enhanced in the current dataset (Fig.~\ref{fig:CFevsFe_APOGEE_Fepoor}). Moreover, s-process elements are recognised not to be particularly enhanced in clump stars in literature analyses beyond some mass transfer binary exceptions \citep[e.g.][]{Merle2016}. Therefore, even though we find evidence for N depletion in clump stars, theoretical understanding remains lacking. \\

\section{Conclusions}
Combining asteroseismology ({\it Kepler}) and survey spectroscopy for very large samples of evolved stars (APOGEE) we provide
an observational determination of the evolution of surface nitrogen abundance as a function of RGB evolutionary state in field stars.  
By selecting samples of thin disk stars at both solar and lower metallicity, we verify the dependence of the first dredge-up on mass and metallicity. We also confirm the universality of extra-mixing along the upper part of the RGB. Furthermore, the data show that there has been significant N depletion between the RGB tip and the He-burning phase/clump evolutionary stages. We propose that mixing with the outer envelope occurs during the He flash, despite the lack of a robust theoretical model of this process. We note that this is the first observational study of sufficient accuracy and size to establish this result. As yet star cluster studies lack enough sufficiently precise N abundances along the RGB and red clump (i.e. better than 0.2 dex) to confirm this result. Our result may provide some interesting clues about the CN and CH band strength paradox as observed in clusters, e. g. \citet{Martell2011}, as well as the puzzle of Li-rich clump stars \citep{Kumar2011}. \\
Note: after this work was submitted, the DR13 APOGEE data have been released. We checked and we can confirm that the main results of this paper remain unchanged with those new data.

\section*{Acknowledgments}
We thank M. Salaris for providing the BaSTI yields.
This work was partly supported by
the European Union FP7 programme through ERC grant
number 320360.
NL acknowledges financial support from the Marie Curie Intra-European fellowship (FP7-PEOPLE-2012-IEF) and the CNES postdoctoral fellowship 2016.
AM and YE acknowledge the support of the UK Science and Technology 
Facilities Council (STFC).
Funding for the Stellar Astrophysics Centre (SAC) is provided by The 
Danish National Research Foundation (Grant agreement no.: DNRF106)
\bibliographystyle{aa}
\bibliography{APOKASC}

\label{lastpage}

\end{document}